\documentclass[english]{article}
\usepackage[T1]{fontenc}
\usepackage[latin1]{inputenc}
\usepackage{graphicx}

\makeatletter

\usepackage{geometry}



\usepackage{babel}
\makeatother

\begin{document}

\title{{\huge The shadow of light: evidences of photon behaviour contradicting
known electrodynamics}{\normalsize \bigskip{}
 }}

\author{Fabio Cardone$^{1,2}${\normalsize , Roberto Mignani$^{2,3}$, {}
Walter Perconti$^{1}$, Andrea Petrucci$^{1}$and Renato Scrimaglio$\bigskip^{4}$}\\
{\normalsize{} {} 
$^{1}$Istituto per lo Studio dei Materiali Nanostrutturati (ISMN
-- CNR)}\\
{\normalsize{} {} Via dei Taurini - 00185 Roma, Italy}\\
{\normalsize{} {} $^{2}$GNFM, Istituto Nazionale di Alta Matematica
\char`\"{}F.Severi\char`\"{}}\\
{\normalsize{} {} \ Città Universitaria, P.le A.Moro 2 - 00185 Roma,
Italy}\\
{\normalsize{} {} $^{3}$Dipartimento di Fisica \textquotedblright
E.Amaldi\textquotedblright , Università degli Studi \textquotedblright
Roma Tre\textquotedblright }\\
{\normalsize{} {} \ Via della Vasca Navale, 84 - 00146 Roma, Italy}\\
{\normalsize{} {} $^{4}$Dipartimento di Fisica, Università degli
Studi de L'Aquila}\\
{\normalsize{} {} via Vetoio 1 - 67010 Coppito, L'Aquila , Italy}}

\maketitle
\begin{abstract}
We report the results of a double-slit-like experiment in the infrared
range, which evidence an anomalous behaviour of photon systems under
particular (energy and space) constraints. The statistical analysis
of these outcomes (independently confirmed by crossing photon beam
experiments in both the optical and the microwave range) shows a significant
departure from the predictions of both classical and quantum electrodynamics. 
\end{abstract}

\section{Introduction}

In the last years we carried out two optical experiments of the double-slit
type, aimed at searching for a possible anomalous photon behavior,
which provided strong clues for a discrepancy with the predictions
of classical and/or quantum electrodynamic$^{(1-4)}$. They originated
from an analysis$^{(5,6)}$ of the Cologne$^{(7)}$ and Florence$^{(8)}$
microwave experiments, which evidenced propagation of electromagnetic
evanescent waves at superluminal speed. Superluminality is naturally
associated to a breakdown of local Lorentz invariance (LLI), and therefore
to a possible anomalous photon behavior. If evanescent waves are identified
with virtual photons$^{(9)}$, such anomalies are expected to occur
within a length scale of the order of the near field size. The analysis
of ref.{[}5] showed that the electromagnetic breakdown of LLI related
to superluminal propagation of evanescent waves exhibits a threshold
behavior both in energy ($E_{0,em}\simeq$4.5$\mu V$) and in space
($\ell\simeq$9\emph{\ }$cm$) (in the sense that it is expected
to occur at energies and distances \emph{lower} than the threshold
values)%
\footnote{More details about the connection between superluminality and LLI
breakdown can be found in refs.{[}5,6].%
}. A repetition of those interference-like experiments with a large
statistics in 2005-2006 allowed us to get a definite evidence for
a photon behavior contradicting standard electrodynamics, under suitable
space and energy constraints.

\section{Experimental setup}

All the experiments were carried out at the microelectronics laboratory
of L'Aquila University.\textbf{\ }The apparatus employed (schematically
depicted in Fig.1) consisted of a Plexiglas box with wooden base and
lid.

The box (thoroughly screened from those frequencies susceptible of
affecting the measurements) contained two identical infrared (IR)
LEDs, as (incoherent) sources of light, and three identical detectors
(A, B, C). In all experiments the LEDS were of the kind High Speed
Infrared Emitter AlGaAs (HIRL 5010, Hero Electronics Ltd.), with emission
peak at 850 $nm$\ and angular aperture of 20$^{\circ}$.The two
sources S$_{1}$, S$_{2}$ were placed in front of a screen with three
circular apertures F$_{1}$, F$_{2}$, F$_{3}$ on it. The apertures
F$_{1}$ and F$_{3}$ were lined up with the two LEDs A and C respectively,
so that each IR beam propagated perpendicularly through each of them.
The geometry of this equipment was designed so that no photon could
pass through aperture F$_{2}$ on the screen. Let us stress that the
apparatus was sized according to the analysis$^{(5)}$ of the superluminal
propagation experiments {[}7,8]. In particular, the dotted line \textit{S}
in Fig.1 corresponds to the horizontal distance between the planes
of the horn antennas in the Florence experiment%
\footnote{In this connection, let us notice that the dotted line \textit{S}
in Fig.1 is a mere geometrical one, and does not represent any physical
trajectory of photons emitted by the source S$_{2}$, since the aperture
F$_{2}$ was well outside the emission cone of S$_{2}$.%
}.

\begin{figure}
\begin{centering}
\includegraphics{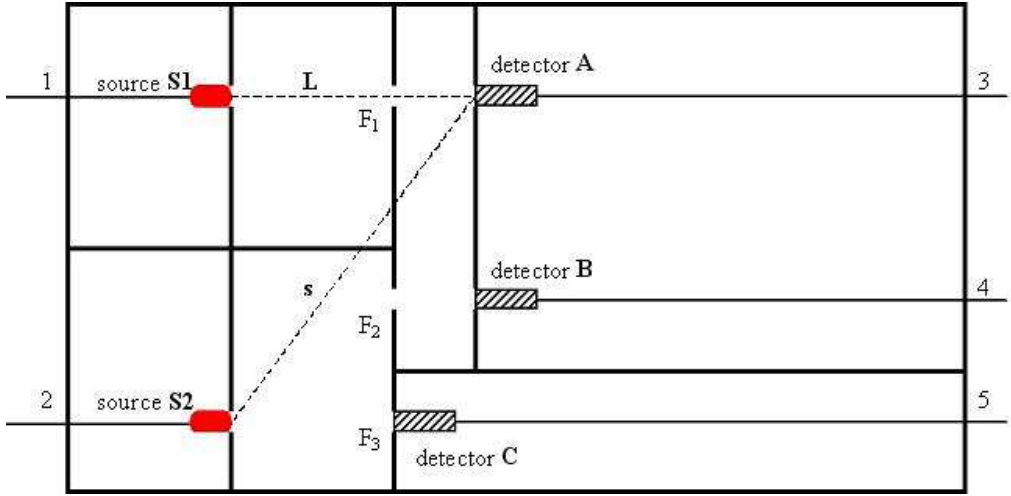} 
\par\end{centering}

\caption{\textit{Above view of the experimental apparatus used in the first
double-slit experiment.}}

\end{figure}

The wavelength of the two photon sources was $\lambda$ = 8.5$\times$10$^{-5}$
$cm$. The apertures were circular, with a diameter of 0.5 $cm$,
much larger than $\lambda$. We worked therefore in absence of single-slit
(Fresnel) diffraction. However, the Fraunhofer diffraction was still
present, and its effects have been taken into account in the background
measurement.

Detector C was fixed in front of the source S$_{2}$; detectors A
and B were placed on a common vertical panel (see Fig.1).

Let us highlight the role played by the three detectors. Detector
C destroyed the eigenstates of the photons emitted by S$_{2}$. Detector
B ensured that no photon passed through the aperture F$_{2}$. Finally,
detector A measured the photon signal from the source S$_{1}$.

In summary, detectors B and C played a controlling role and ensured
that no spurious and instrumental effects could be mistaken for the
anomalous effect which had to be revealed on detector A. The design
of the box and the measurement procedure were conceived so that detector
A was not influenced by the source S$_{2}$ according to the known
and officially accepted laws of physics governing electromagnetic
phenomena: classical and/or quantum electrodynamics. In other words,
with regards to detector A, all went as if the source S$_{2}$ would
not be there at all or be always kept turned off.

In essence, the experiments just consisted in the measurement of the
signal of detector A (aligned with the source S$_{1}$) in two different
states of source lighting. Precisely, a single measurement on detector
A consisted of two steps:\medskip{}

1)\qquad{}Sampling measurement of the signal on A with source S$_{1}$
switched on and source S$_{2}$ off;

2)\qquad{}Sampling measurement of the signal on A with both sources
S$_{1}$ and S$_{2}$ on.\medskip{}

As already stressed, due to the geometry of the apparatus, no difference
in signal on A between these two source states ought to be observed,
according to either classical or quantum electrodynamics.\emph{\ }If
$A($S$_{1}i$ S$_{2}k)$ ($i,k=on,off$) denotes the value of the
signal on A when source S$_{1}$ is in the lighting state $i$ and
S$_{2}$ in the state $k$, a possible non-zero difference $\Delta A=A($S$_{1}on$
S$_{2}off)-A($S$_{1}on$ S$_{2}on)$ in the signal measured by A
when source S$_{2}$ was off or on (and the signal in B was strictly
null) has to be considered evidence for the searched anomalous effect.

Let us explicitly notice that the geometry of the box was critical
in order to reveal the anomalous photon behavior.

\section{The first two experiments}

The main difference between the first two experiments was in the nature
of the detectors A,\ B, C, which were photodiodes in the former case$^{(1,2)}$
and phototransistors (of the type with a convergent lens) in the latter$^{(3,4)}$
(see refs.{[}1-4] for technical details)\textbf{.} Moreover, in the
second experiment a right-to-left inversion was made along the bigger
side of the box. Thus, it was possible to study how the phenomenon
changes under a spatial parity inversion and for a different type
of detector. In the first experiment the plane containing the detectors
A, B was movable (the distance was varied by steps of 1 $cm$ on the
whole range of 10 $cm$). This allowed us to study how the phenomenon
changes with distance from the sources.

The outcomes of the first experiment were positive, namely \emph{the
differences} $\Delta A$ \emph{between the measured signals on detector
A in the two conditions were different from zero.} Moreover, the phenomenon
obeyed the threshold behavior predicted by the analysis {[}5] of the
Cologne and Florence experiments. In particular, $\Delta A$ ranged
from (2.2$\pm$0.4)$\mu V$ to (2.3$\pm$0.5)$\mu V$, values well
below the threshold energy $E_{0,em}$= 4.5$\mu V$ , and the anomalous
effect was observed within a distance of at most 4 $cm$ from the
sources. The dependence of the phenomenon on the detector-source distance
highlights further the critical role of the geometry of the box in
the detection of the phenomenon.

We can consider such an effect as the consequence of \emph{a \textquotedblright
hidden\textquotedblright\ (or virtual) interference between the
photon beams of the two sources}. This must be meant in the sense
that something like a \char`\"{}virtual screening\char`\"{} occurred,
which modified the photon-photon cross section thus producing a change
in the number of photons detected by A (\emph{\char`\"{}shadow of
light\char`\"{}})%
\footnote{A more detailed discussion of this \char`\"{}shadow of light! and
its possible interpretation can be found in refs.{[}2-4].%
}.

The results of the second experiment confirmed those of the first
one. The value of the difference measured on detector A was (0.008$\pm$0.003)$\mu V$,
which is consistent, within the error, with the difference $\Delta A$
$\simeq$ 2.3 $\mu V$ measured in the first experiment, \emph{provided
that the unlike efficiencies of the phototransistors with respect
to those of the photodiodes are taken into account}.%
\footnote{One can define the relative geometrical efficiency $\eta_{g}$ of
the phototransistor (with respect to the photodiode) as the ratio
of their respective sensitive areas, and their relative time efficiency
$\eta_{t}$ as the ratio of their respective detection times. Then,
one can define the relative total efficiency $\eta_{T}$ of the phototransistor
with respect to the photodiode as the product $\eta_{T}$=$\eta_{g}\eta_{t}$
. From the values of $\eta_{g}$ and $\eta_{t}$ in this case, one
gets$^{(3)}$ $\eta_{T}$=0.0015. Therefore, it was reasonable to
foresee that the value of the expected phenomenon in the second experiment
to be given by the product of the total relative efficiency times
the value measured in the first experiment, \textit{i.e.} $\eta_{T}\left[(2.3\pm0.5)\mu V\right]$=$\left(0.004\pm0.001\right)\mu V$,
in agreement with the experimental result.%
}

The consistency between the results of the first two experiments shows
apparently that the effect is not affected by the parity of the equipment
and by the type of detector used (at least for photodiodes and phototransistors).

Furthermore, a different time procedure to sample the signals on the
detectors was used in the two experiments. We indeed realized that
the sampling time procedure was apparently crucial in order to observe
the anomalous interference effect. This is due to the fact that the
phenomenon has a peculiar time structure that makes the sampling procedure
critical$^{(3)}$.\ Therefore, in order to optimize the performance
of the effect detection for different detectors, statistics being
equal, it was necessary to suitably change the time sampling. This
latter consisted of two time steps, namely the waiting time $t_{w}$
(defined as the time interval between the lighting of the source(s)
and the start of the sampling on the detectors), and the measurement
time $t_{m}$, \textit{i.e.} the actual interval during which measurements
were taken. In the first experiment, it was $t_{w}$=60 $s$ and $t_{m}$
was determined manually, whereas in the second one it was $t_{w}=1s$
and $t_{m}$\ =5 $s$. Then, it turned out that there was apparently
a sort of unavoidable bond between detector and sampling-time procedure,
to be taken into account in order to reveal the effect.

We stress that the results of the double-slit experiments have been
independently supported by experiments with orthogonal crossing photon
beams, in which similar anomalous effects have been observed. We refer
to two interference experiments (carried out after our first one),
one with microwaves emitted by horn antennas, at IFAC - CNR (Ranfagni
and coworkers)$^{(10,11)}$, and the other with infrared CO$_{2}$
laser beams, at INOA-CNR (Meucci and coworkers)$^{(3,12)}$.

\section{Third experiment}

The third experiment was planned and carried out in order to obtain
a further evidence of the observed effect, by clarifying some of its
features. In order to test the apparent bond between detectors and
sampling time procedures, the experiment was carried out by means
of the box with photodiodes but using the sampling-time procedure
adopted with phototransistors (namely $t_{w}$= 1 $s$\ and $t_{m}$\ =
5 $s$). Our aim was just to put ourselves in the worst possible situation
with respect to the effect detection.

Let us note that the photodiodes used as detectors in the first and
third experiment were integrated to a transimpedance amplifier type
OPT301 of Burr-Brown (registered Trade Mark), transducing the photocurrent
signal into a voltage signal. Such a voltage, measured by means of
a multimeter 34401A of Agilent, did not depend therefore on the value
of the circuit resistances of the voltage measuring system.

As we shall see, the results of this third experiment were consistent
with those of the two previous ones. Moreover, the measurements were
repeated several times over a whole period of four months, in order
to collect a fairly large amount of samples and hence have a significant
statistical reproducibility of the results. Thanks to this large quantity
of data, it was possible to study the distribution of the differences
of signals on detector A, which is shown in Fig. 2. For clarity's
sake, we reported only the differences $\Delta A$ outside the interval
$\left[-1,1\right]$, which is the interval of compatibility with
zero of the values of $\Delta A$. The data of the single measurements
have been suitably treated in order to get rid of the instrumental
drift.

\begin{figure}
\begin{centering}
\includegraphics{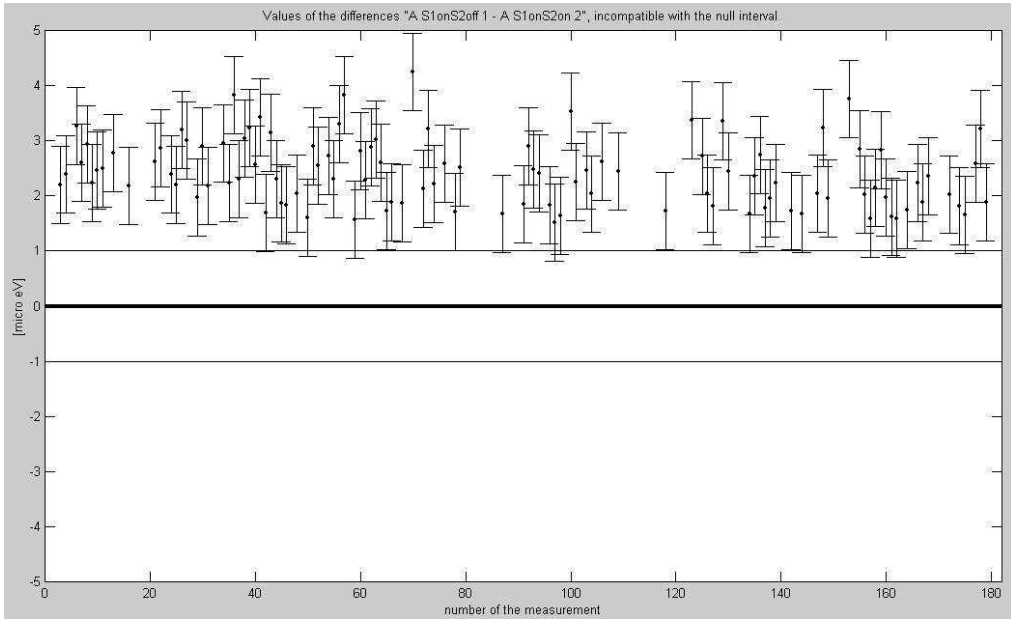} 
\par\end{centering}

\caption{\textit{Value of the differences} $\Delta A$ \textit{of signal sampled
on detector A for the two lighting states of the sources S}$_{1}$\textit{\ on,
S}$_{2}$\textit{\ off, and S}$_{1}$\textit{\ on, S}$_{2}$\textit{\ on
(third experiment). The differences are clearly incompatible with
zero.}}

\end{figure}

We want now to show that a more detailed analysis of the measurements
of the third experiment are just in favour of the anomalous interference
observed as signature of a possible violation of electrodynamics.

This is easy to realize, by noting that the distribution of the results
of the third experiment (reported in Fig.2) is unmistakably different
from that expected from the theoretical predictions of both quantum
and classical electrodynamics.

\begin{figure}
\begin{centering}
\includegraphics{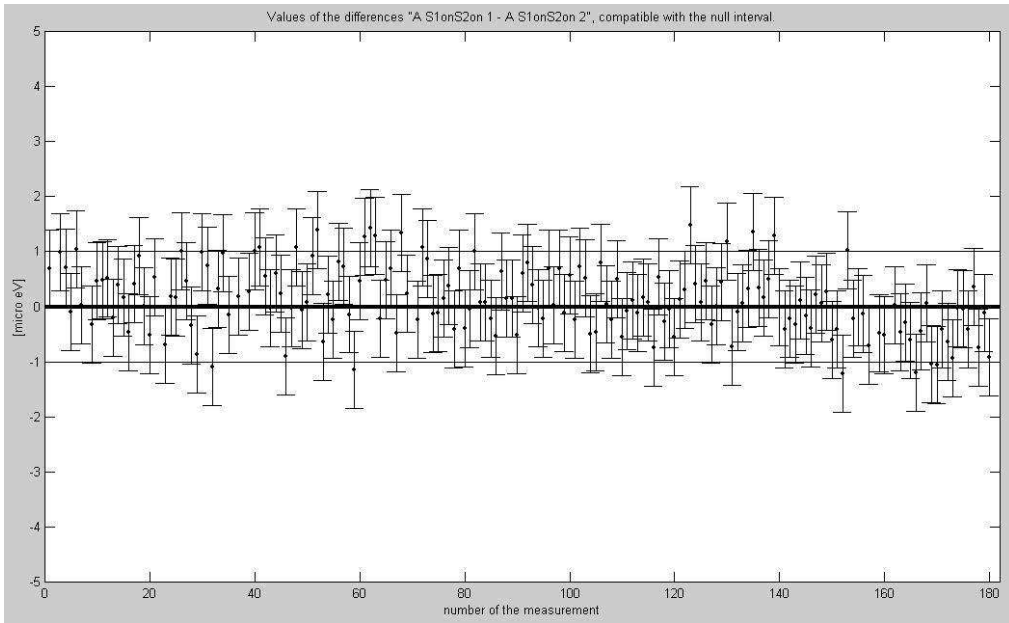} 
\par\end{centering}

\caption{\textit{Values of the differences} $\Delta A^{\prime}$ \textit{of
signal sampled on detector A with both sources on (third experiment).
The differences are clearly compatible with zero.}}

\end{figure}

In this connection, let us recall that Fig.2 shows the signal differences
measured on A in correspondence to the two \emph{different} states
of lighting of the source S$_{2}$, $\Delta A=A($S$_{1}on$ S$_{2}off)-A($S$_{1}on$
S$_{2}on)$. For comparison, we report in Fig.3\ the differences
of the two values sampled on A in the \emph{same} lighting condition
of the sources,\textit{\ i.e.} with both sources turned on: $\Delta A^{\prime}=A($S$_{1}on$
S$_{2}on)-A($S$_{1}on$ S$_{2}on)$. For clarity's sake, we show
only the differences inside the null interval $\left[-1,1\right]$.
There is no surprise in observing that the differences $\Delta A^{\prime}$
are almost evenly distributed around zero, since the subtracted values
belong to the same population. However, by the very design of the
experimental box, according to either classical or quantum electrodynamics
detector A had not to be affected by the state of lighting of the
source S$_{2}$. Hence, one would expect that the mean value of the
differences $\Delta A$ (corresponding to the two different lighting
states of the source S$_{2}$) was zero and that these differences
were uniformly distributed around it. In other words, one would expect
to find roughly the same number of positive and negative differences,
and therefore that both Fig.2 and Fig.3 displayed two compatible distributions
of differences evenly scattered across zero. On the contrary, \emph{the
differences} $\Delta A$ \emph{in Fig.2 are not uniformly distributed
around zero but are markedly shifted upward} (as compared to those
in Fig. 3), and hence the number of positive differences is larger
than the negative ones. This upward shift means that $A($S$_{1}on$
S$_{2}off)>A($S$_{1}on$ S$_{2}on)$, and hence that the signal on
detector A is lower when both of the incoherent sources are on%
\footnote{Actually, due to the very operation of the used photodiodes, detector
A measured a lower number of photons when the number of photons in
the box was higher.%
}. Of course, the incoherence of sources excludes the possibility that
the signal lowering could be due to destructive interference. We can
conclude that \emph{distribution 3 is compatible with zero (as it
must be), whereas distribution 2 is not, at variance with the predictions
of either classical and quantum electrodynamics.}

In order to further enforce the evidence for the difference of the
two physical situations corresponding to Figs.2 and 3, we carried
out a statistical analysis of the results found in the two cases (only
the differences outside the interval $\left[-1,1\right]$ have been
considered), by taking into account the instrumental drift. The Gaussian
curves obtained are shown in Fig.4. The dashed, red curve refers to
the signal differences $\Delta A=A($S$_{1}on$ S$_{2}off)-A($S$_{1}on$
S$_{2}on)$, whereas the solid, blue one to $\Delta A^{\prime}=A($S$_{1}on$
S$_{2}on)-A($S$_{1}on$ S$_{2}on)$. The two curves differ by 3.82
$\sigma$, clearly showing that the two cases are statistically distinct,
the latter one representing a mere fluctuation (unlike the former).
Furthermore, the mean value corresponding to the Gaussian of the differences
$\Delta A$ is $\overline{\Delta A}$ = 2.41 $\mu eV$, a value in
full agreement with the results of the first two experiments$^{(1-4)}$.

\begin{figure}
\begin{centering}
\includegraphics[height=8cm]{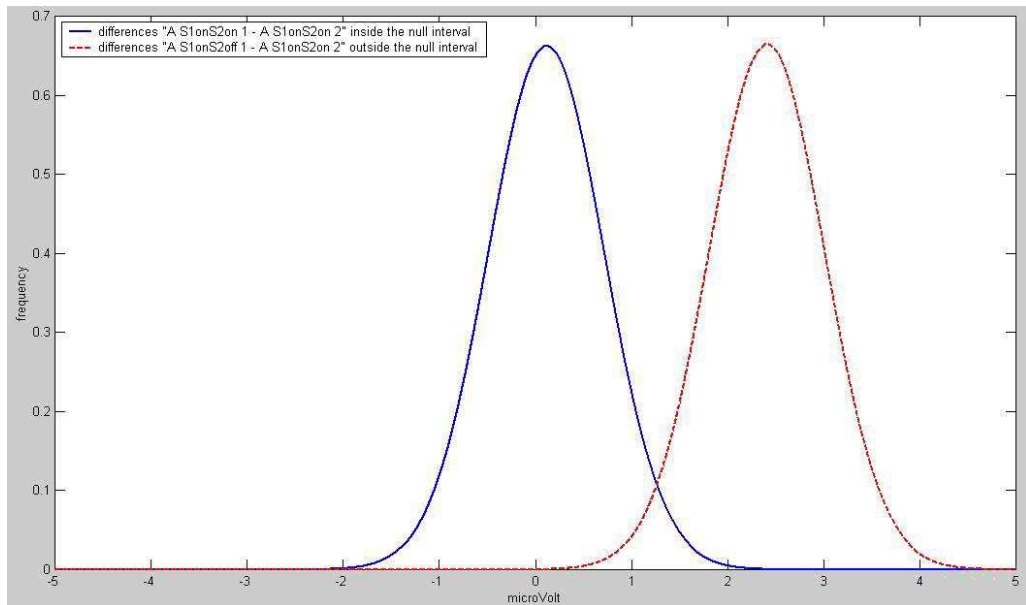} 
\par\end{centering}

\caption{\textit{Gaussian curves (normal frequency vs. signal difference in}
$\mu V$) \textit{for the signal differences} $\Delta A$ \textit{and}
$\Delta A^{\prime}$\textit{on detector A for the two cases of source
S}$_{2}$ \textit{off and on (dashed and solid curve, respectively).
The instrumental drift has been taken into account. It is} $\overline{\Delta A}$=\textit{2.411}
\textit{(}$\sigma$\textit{=0.601);} $\overline{\Delta A^{\prime}}$\textit{=0.116
(}$\sigma$\textit{=0.602).}}

\end{figure}

As a further check, we carried out an analysis of the data of the
third experiment by constraining the instrumental drift to be a constant.
As is well known, such a procedure makes the observed effect to disappear
if it is a mere instrumental one. On the contrary, such a strong constraint
did not affect the set of data, which remains statistically significant.
This means that \emph{the observed anomalous interference has not
an instrumental origin.}

We can therefore conclude that the results obtained on the anomalous
behavior of photon systems --- apparently at variance with usual (classical
and quantum) electrodynamics --- bring to light a more complex physics
of the electromagnetic interaction, which calls for a critical reexamination
of standard electrodynamics and quantum mechanics$^{(4)}$.

\textbf{Acknowledgements -} Useful discussions with A. Ranfagni are
gratefully acknowledged. Moreover, a special thank is due with sincere
pleasure to the President of CNR Fabio Pistella, who, not only now
but even previously in his capacity of President of INOA (National
Institute of Applied Optics), has encouraged and supported the execution
of these experiments, and has been actively involved in the discussions
concerning the experimental results.

\newpage{}

\end{document}